 \documentclass[final,3p,times,twocolumn,sort&compress]{elsarticle}

 \usepackage{graphicx,amssymb,amsthm,lineno}
 \usepackage{dcolumn,multirow}

\biboptions{square}

\journal{Brazilian Journal of Physics}

\begin{document}

\begin{frontmatter}



\title{Kauffman cellular automata on quasicrystal topology}


\author[ufsa]{Carlos Handrey Araujo Ferraz\corref{cor1}}
\ead{handrey@ufersa.edu.br}
\author[ufsa]{Jos\'e Luiz Sousa Lima}
\ead{jlima@ufersa.edu.br}
\cortext[cor1]{Corresponding author}
\address[ufsa]{Exact and Natural Sciences Center, Universidade Federal Rural do Semi-\'Arido-UFERSA, PO Box 0137, CEP 59625-900, Mossor\'o, RN, Brazil}

\begin{abstract}
In this paper we perform numerical simulations to study Kauffman cellular automata (KCA) on quasiperiod lattices. In particular, we investigate phase transition, magnetic entropy and propagation speed of the damage on these lattices. Both the critical threshold parameter $p_{c}$ and the critical exponents are estimated with good precision. In order to investigate the increase of statistical fluctuations and the onset of chaos in the critical region of the model, we have also defined a magnetic entropy to these systems. It is seen that the magnetic entropy behaves in a different way when one passes from the frozen regime ($p<p_{c}$) to the chaotic regime ($p>p_{c}$). For a further analysis, the robustness of the propagation of failures is checked by introducing a quenched site dilution probability $q$ on the lattices. It is seen that the damage spreading is quite sensitive when a small fraction of the lattice sites are disconnected. A finite-size scaling analysis is employed to estimate the critical exponents. From these numerical estimates, we claim that on both pure ($q=0$) and diluted ($q=0.05$) quasiperiodic lattices, the KCA model belongs to the same universality class than on square lattices. Furthermore, with the aim of comparing the dynamical behavior between periodic and quasiperiodic systems, the propagation speed of the damage is also calculated for the square lattice assuming the same conditions.  It is found that on square lattices the propagation speed of the damage obeys a power law as $v\sim (p-p_{c})^{\alpha}$, whereas on quasiperiod lattices it follows a logarithmic law as $v \sim \ln(p-p_{c})^\alpha$.
\end{abstract}


\end{frontmatter}


\section{Introduction}

Kauffman cellular automata (KCA)~\cite{kauffman69} or more generally random Boolean networks have been studied in the past to describe genetic regulatory networks but due to their general features since they do not assume any particular function of the nodes these can also be used to study a large variety of important issues concerning sychronization~\cite{morelli2001}, stability~\cite{bilke2002}, robustness~\cite{reka2004} and control of chaos~\cite{luque97}, just to mention a few examples. In particular, we are interested in studying how small failures (damages) produced on complex structures of automata propagate throughout the entire system. These failures may stand for genetic mutations, fractures, infectious disease spreading and virus propagation on computer networks.  Earlier studies related to failure propagation on complex structures of automata have focused either on periodic lattices~\cite{derrida86,stauffer87,arcangelis,stauffer93}, which present short-range interactions, or on random graphs~\cite{pomeau86,ferraz2007} in which long-range interactions take place. However, there have been few studies addressing the dynamics of the failure propagation in systems that possess short-range interactions with breaking of translational symmetry. The lack of this symmetry could influence the propagation of the damage cloud as well as change both the critical threshold parameter $p_{c}$ and the universality class of these systems. The quasicrystals' topology is quite suitable for such a study since it does not present neither periodic translational nor close orientational order. In fact, such systems can exhibit rotational symmetries otherwise forbidden to crystals~\cite{levine86}. Furthermore, given the lack of periodicity of these systems, only numerical approaches can be performed. Under a point of view purely geometrical, quasicrytals can be thought as quasiperiodic lattices. Recently, a monte carlo study~\cite{ferraz2015} has confirmed that both the periodic lattices and the quasiperiodic lattices belong to the same universality class~\cite{stanley}, despite the critical temperature of these lattices being different. It is the purpose of this paper to present some numerical results concerning both criticality and the dynamics of the failure propagation on such quasiperiodic lattices and thereby understanding how this topology can influence the time evolution of these systems.

\begin{figure}[!t]
 \centering
 \includegraphics*[scale=0.35,angle=-90]{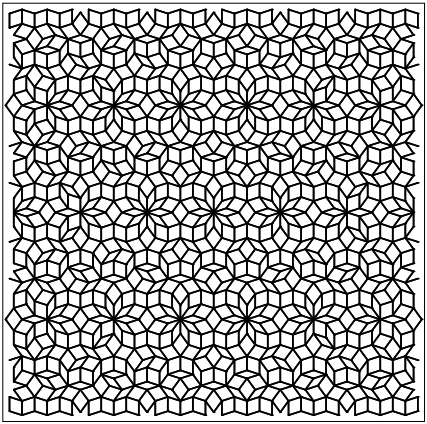}
 \caption{\label{fig:1} Quasiperiodic lattice generated by the strip projection method~\cite{ferraz2015}. The lattice is shown inside a square projection window. The periodic boundary conditions are imposed at the lattice sites closer to the projection window.}
\end{figure}

The phase transition of the KCA model is defined by calculating the Hamming distance (see section \ref{sec:cp}) between almost identical lattices (only a small number of sites have different states). We denote this Hamming distance as the damage; if it remains localized or eventually vanishes, then one says that the system is in a frozen phase. Otherwise, if the damage spreads out over a considerable part of the system, then one says that the system is in a chaotic phase. Usually, the control parameter of the model~\cite{stauffer87, ferraz2007, ferraz2008} is the probability $p$ for the boolean function, which rules the time evolution of a given site, yields as output the value $1$. Following Ref.~\cite{luque97}, here `chaos' is not the usual low-dimensional deterministic chaos but a phase where damage spreading takes place.

In this paper we perform numerical simulations to study KCA on quasiperiodic lattices. In particular, we investigate phase transition, magnetic entropy and propagation speed of the damage~\cite{ferraz2008} on these lattices. Both the critical threshold parameter $p_{c}$ and the critical exponents were estimated for different treated cases.

In order to investigate the increase of the statistical fluctuations and the onset of chaos in the critical region of the model, we have also defined a magnetic entropy to these systems. It is seen that the magnetic entropy behaves in a different way when the system changes from the frozen regime ($p<p_{c}$) to the chaotic regime ($p>p_{c}$). For a further analysis, the robustness of the propagation of failures~\cite{ferraz2008} was checked by introducing a quenched site dilution probability $q$ on the lattices. It is found that the damage spreading is quite sensitive when a small fraction of sites are disconnected from these lattices. The quasiperiodic lattices analyzed here were generated using the strip projection method~\cite{ferraz2015} with each automaton placed in the vertices of the rhombi that make up the lattice (Fig.~\ref{fig:1}). For this type of lattice, the number of nearest neighbors at a given site can vary from $K=3$ to $K=10$ with a mean coordination number equal to $<z>=3.98$. A finite-size scaling analysis was used to estimate the critical exponents and periodic boundary conditions were imposed on the generated lattices in order to reduce finite-size effect in the simulations.  We calculate several quantities including the order parameter, logarithmic derivative of the order parameter, propagation speed of the damage and magnetic entropy. Moreover, with the aim of comparing the dynamical behavior between periodic and quasiperiodic systems, the propagation speed of the damage was also calculated for square lattices assuming the same conditions.

This paper is organized as follows.  In section \ref{sec:kaf}, we give a brief review about the Kauffman model. Next, we describe the computational
procedure used to implement the model on quasiperiodic lattices as well as to accomplish a site dilution on these lattices. After that, in section \ref{sec:mpy}, we define the absolute magnetic entropy on KCA. In section \ref{sec:r},  we present our results concerning the phase transition, entropy and speed propagation of the damage. In section \ref{sec:c}, we conclude by summarizing the main results and providing recommendations for further research.

\section{\label{sec:kaf} The Kauffman Model}

Kauffman originally introduced networks of Boolean automata in order to study the behaviour of generic
regulatory systems. The basic idea of the Kauffman model is to consider a mixture of all possible binary cellular automata. The
Kauffman model can be realized on a lattice, by choosing Boolean rules individually for each site.  Each of $N$ lattice sites hosts a
Boolean variable  $\sigma_{i}$ (spin up or down) which is either zero or unity. The time evolution of this model is determined by $N$
functions $f_{i}$ (rules) which are randomly chosen for each site independently, and by the choice of $K$ input sites \{$j_{K}(i)$\}
for each site $i$. Thus the value $\sigma_{i}$ at site $i$ for time $t$+1 is given by:

\begin{equation}\label{eq:1}
\sigma_{i}(t+1)=f_{i}(\sigma_{j_{1}}(t), \ldots,\sigma_{j_{K}}(t))\quad i=1,2,\ldots,N.
\end{equation}

Each Boolean function $f_{i}$ is specified, once a value is given for each one of the $2^{K}$ possible neighbour configurations. A variable $\sigma_{i}$ is called
relevant for the spreading damage process if it is unstable i.e. the state of  other variables \{$\sigma_{j}$\} depend on
$\sigma_{i}$.  If one imposes that the inputs and the chosen Boolean functions do not change with time, we have the quenched
Kauffman model. On the other hand, if one admits that both change with time, we have the annealed Kauffman model. A big difference
between the two cases is that in the quenched case there are limit cycles and in the annealed case not. Here we consider only the quenched case.
In this case, as the time development is totally deterministic, and since $N$ different Boolean variables can produce $2^{N}$ different lattice configurations, we must return after at most $2^{N}$ time-steps to the previous initial configuration. Then the system will repeat the same configurations, staying within this limit cycle. For the nearest-neighbour Kauffman model on the square lattice, the number of relevant limit cycles increases exponentially with system size in the non-chaotic phase~\cite{flyvbjerg86}. Kauffman identified these different limit cycles with the different cell types in our body and found that their number grows as $\sqrt{N}$ for $N$ interacting genes. The annealed case can be solved analytically, whereas for the quenched case only computer simulations were performed up to now.

\section{\label{sec:cp} Computational Procedure}

A standard way to implement the Kauffman model is introducing a parameter $p$ such that for each site on the lattice we select among the ${{2}^2}^K$ rules one which for each outcome will have  spin up with probability $p$. In a computer simulation, first one goes through all $N$ sites of the system, and for each site one goes through all ${2}^K$ neighbour configurations, and for each such configuration one determines by drawing a random number if its spin will be up or down; if the random number is smaller than $p$ then its spin will be up, otherwise it will be down. Once one has gone through all neighbour configurations of that site, then one has fixed the rule for that site, and one can go to the next site. After that one selects an initial configuration of the Boolean variables by randomly assigning to each lattice site a spin up or down with equal probability.  We will consider two systems (replicas), identical in the connections and rules, and also identical in the initial configuration of the Boolean variables, except that on one of them we flip the most central sites of the lattice (around 0.5\% of the lattice sites) at every time-step along the simulation.  The number of spins which at time $t$ is different between the two replicas is called the Hamming distance $d(t)$  or simply the failure. For two lattice configurations \{$\sigma_{i}(t)$\} and \{$\rho_{i}(t)$\}, we have

\begin{equation}\label{eq:2}
d(t)=\frac{1}{N}\sum_{i}|\sigma_{i}(t)-\rho_{i}(t)|,
\end{equation}
and we can define an order parameter $\psi$ for the system taking in Eq.~\ref{eq:2} the limit $t\rightarrow \infty$, namely

\begin{equation}\label{eq:3}
\Psi=\lim_{{d(0)\rightarrow 0}}d(\infty).
\end{equation}
Computationally, convergence is typically reached after a few thousand  time steps. In this way we can study the phase transition, entropy and the propagation speed of the damage cloud varying the value $p$. It has been observed that $\Psi$ goes to zero at the  critical threshold parameter $p_{c}$ in systems with dimensions greater than one for the short-range case of the Kauffman model, in a similar way to the para-ferromagnetic phase transition. In other words, for all $p \leqslant p_{c}$,  a small initial damage vanishes or remains small, i.e. it belongs to a small cluster of `damaged spins' after a sufficiently long time. One says that the system is in the frozen phase. On the other hand, for all $p>p_{c}$, a small initial damage spreads throughout a considerable part of the system. Then one says that the system is in the chaotic phase. Of particular interest is, however, the border case $p=p_{c}$ where fractal properties appear~\cite{coniglio87}. Obviously, $p$ and $1-p$ are statistically equivalent, so that we do not consider $p>0.5$.

In order to check the robustness of the propagation of the damage cloud, we have also introduced a quenched site dilution probability $q$ on the lattices. Thus we consider the  Kauffman model on quenched site-diluted quasiperiodic lattices. The dilution procedure is as follows: at the beginning of each simulation run, we generate a new configuration lattice starting from a pure lattice ($q=0$) by disconnecting each lattice site with a probability $q\neq0$. After this procedure, we are left with a new lattice with a density $1-q$ of linked sites.

In the transient regime, the propagation speed $v$ required for the damage to spread throughout the entire system was calculated by measuring the time it takes to touch the lattice boundaries. We perform several calculations of the propagation speed of the damage for both square and quasiperiodic lattices assuming the same conditions, it means that the most central sites were flipped at every iteration along the simulation (i.e., persistently disturbed sites). We wait up to $10^6$ lattice sweeps so that the damage cloud could reach the lattice boundaries. Only succeeded runs in which the damage cloud reached the lattice boundaries were considered in the averages. We average over up to 600 independent runs. Different lattice sizes were considered for each value of $p$. Besides, an extrapolation technique was also used to take into account the thermodynamic limit. This extrapolation was achieved by analyzing how the propagation speed of the damage $v$ depends on the reciprocal of the lattice size ($1/N$) when one takes the limit $N\rightarrow\infty$.

\section{\label{sec:mpy} Magnetic Entropy}

A `magnetic' bath can be associated to the system, where the parameter $p$ will be the intensive thermodynamic variable in this case. Defining the hamiltonian of the system as
\begin{equation}\label{eq:4}
H =  - J\sum\limits_{i = 1}^N {\delta (\sigma _i\oplus\rho_i ,0) -} \delta (\sigma _i\oplus\rho_i,1),
\end{equation}
where $J$ is the energetic coupling constant, $\delta$ is the Kronecker delta function, $\oplus$ is the modulo-2 addition of $\sigma _i$ and $\rho_i$ (spin variables of the $i$th site in interacting replicas). From the fluctuation-dissipation theorem, we can calculate the energy dispersion per spin as
\begin{equation}\label{eq:5}
\Omega (p) = \frac{{K^2 }}{N}( <E^2> - <E>^2 ),
\end{equation}
where $K=J/p$ is the reciprocal of the parameter $p$ (here we assume $J=1$), $E=U/N$ is the energy per spin (see Eq.~\ref{eq:4}) and  $<\ast>$ denotes an average value.

Based on statistical mechanics, one can define a energy function associated to the empirical probability of a given configuration $\mu \equiv \{\sigma _i\oplus\rho _i \}$ (again $\oplus$ is the modulo-2 addition) of the system~\cite{stephens}:
\begin{equation}\label{eq:5a}
U_\mu   =  - \ln (P_\mu),
\end{equation}
The most probable states have low energy, while the less probable states correspond to high energies. On the other hand, the probability of the system to be in a certain configuration of its microscopic states is computed from the energy function. So, we can write the canonical partition function $Z(p)$ as
\begin{equation}\label{eq:5b}
Z(p) = \sum\limits_\mu  {\exp \left\{ {-\frac{1}{p}U_\mu} \right\},}
\end{equation}
where $p$ would be to the temperature in a thermodynamic system, here it is our `magnetic' parameter. Using Eqs.~\ref{eq:5a}, we can rewrite Eq.\ref{eq:5b} as
\begin{equation}\label{eq:5c}
Z(p) = \sum\limits_\mu {P_\mu ^{1/p} },
\end{equation}
which allows to define a $p$-dependent probability distribution $P_{\{ p\}}(\mu )$ as
\begin{equation}\label{eq:5d}
P_{\{ p\} } (\mu ) = \frac{1}{{Z(p)}}P_\mu  ^{1/p}.
\end{equation}

Once the canonical partition function $Z(p)$ is defined, the magnetic entropy $S(p)$ ~\cite{wolf,wolf2,stephens} and the energy dispersion $\Omega(p)$~\cite{stephens} can be respectively derived from it, i.e.,
\begin{equation}\label{eq:5e}
S(p) =  -\sum\limits_\mu  {P_{\{ p\} } (\mu )\log(P_{\{ p\} } (\mu ))} ,
\end{equation}
and
\begin{equation}\label{eq:6}
\Omega  = p\frac{{dS}}{{dp}}.
\end{equation}

By integrating the above relation, one can obtain the entropy per spin at the parameter $p$ as
\begin{equation}\label{eq:7}
S(p) = S(p_0 ) + \int\limits_{p_0 }^p {\frac{\Omega}{p}} dp,
\end{equation}
and next assuming the limit $\mathop {\lim }\limits_{p_0\to 0} S(p_0) = 0$, we finally have the absolute entropy at $p$:

\begin{equation}\label{eq:8}
S(p) = \int\limits_0^p {\frac{\Omega }{p}} dp.
\end{equation}
Eq.~\ref{eq:8} above was evaluated here through numerical integration~\cite{recipes96}.

\section{ \label{sec:r} Numerical Results}

\begin{figure}[!t]
 \centering
 \includegraphics*[scale=0.30,angle=0]{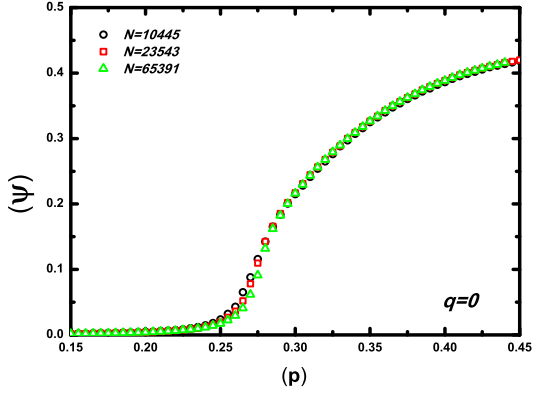}
 \caption{\label{fig:2} Plot of the order parameter $\Psi$ as a function of $p$ for three different lattice sizes for the case $q=0$. One can notice a typical second-order phase transition around $p_{c}=0.27(5)$.}
\end{figure}

\begin{figure}[!t]
 \centering
 \includegraphics*[scale=0.30,angle=0]{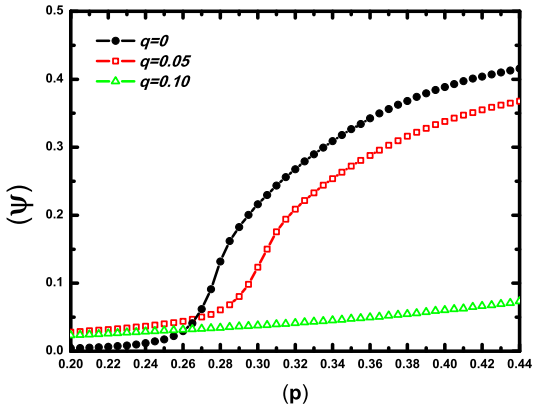}
 \caption{\label{fig:3} Phase transition on a quasiperiodic lattice of size $N=65391$ for three different value of the site dilution rate $q$. For $q=0.10$ there is not a phase transition anymore.}
\end{figure}

\begin{figure}[!t]
 \centering
 \includegraphics*[scale=0.30,angle=0]{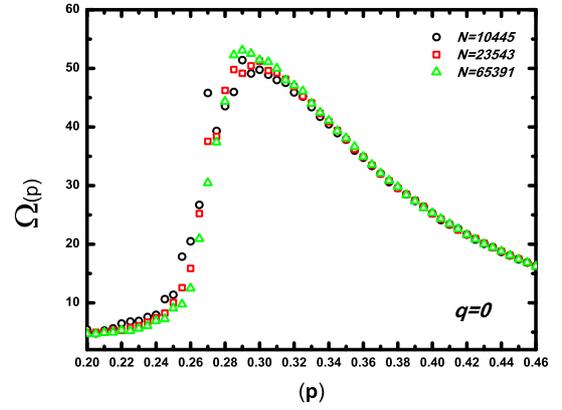}
 \caption{\label{fig:4} Plot of the energy dispersion per spin $\Omega$ as a function of $p$ for three different lattice sizes for the pure case $q=0$. An abrupt increase of the statical fluctuations can be seen around $p_{c}=0.28$.}
\end{figure}

\begin{figure}[!t]
 \centering
 \includegraphics*[scale=0.30,angle=0]{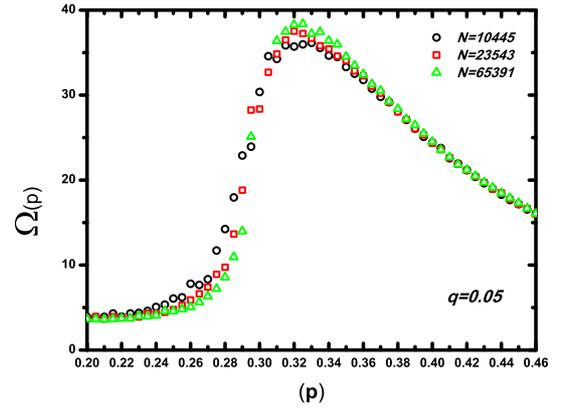}
 \caption{\label{fig:5}  Plot of the energy dispersion per spin $\Omega$ as a function of $p$ for the diluted case $q=0.05$. An abrupt increase of the statical fluctuations can be seen around $p_{c}=0.31$.}
\end{figure}

\begin{figure}[!t]
 \centering
 \includegraphics*[scale=0.30,angle=0]{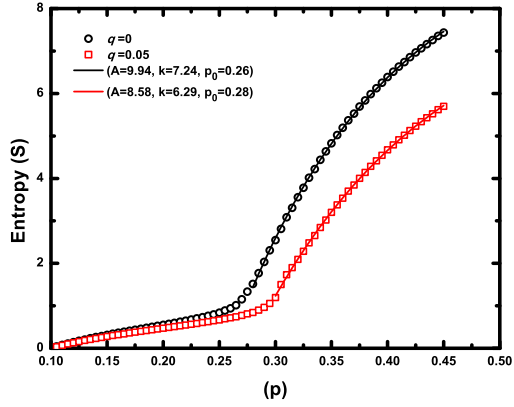}
 \caption{\label{fig:6} The absolute magnetic entropy $S$ as a function of the parameter $p$ considering a system with $N=65391$ sites for both the $q=0$ and $q=0.05$ cases. In the frozen phase, the entropy exhibits a clear linear dependence on $p$ while in the chaotic phase it increases non-linearly as $p$ increases. Lines are the best non-linear fits of the form $S = A(1-\exp (-k(p -p_0 ))$ to the data points above $p_{c}$.}
\end{figure}

\begin{figure}[!t]
 \centering
 \includegraphics*[scale=0.30,angle=0]{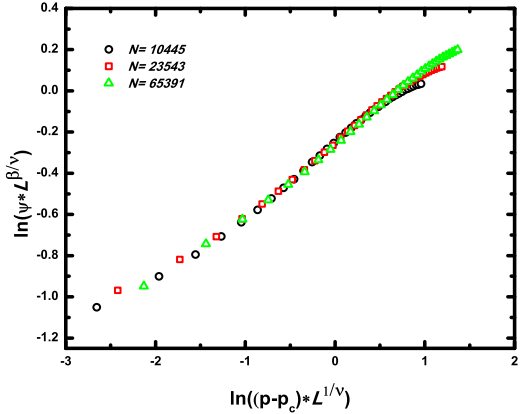}
 \caption{\label{fig:7} Log-log plot of $\Psi L^{\beta/ \nu}$ versus ($p-p_{c}$)$ L^{1/\nu}$ for the case $q=0$.}
\end{figure}

\begin{figure}[!t]
 \centering
 \includegraphics*[scale=0.30,angle=0]{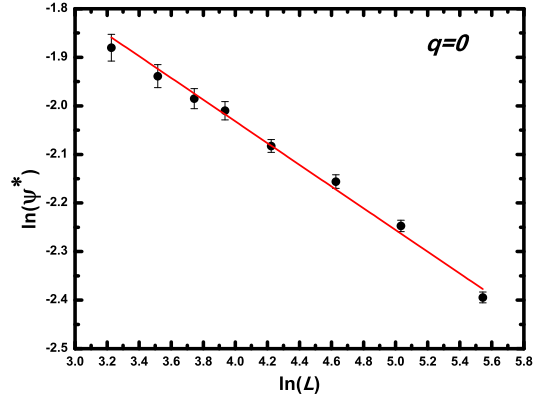}
 \caption{\label{fig:8} Log-log plot of $\Psi^{\ast}$ versus the linear size $L$ of the system for the case $q=0$. The red straight line is the best linear fit to the data ($\chi_{r}^2=1.14$ and a goodness-of-fit probability $Q(\chi_{r}^2)=33.4\%$).}
\end{figure}

\begin{figure}[!t]
 \centering
 \includegraphics*[scale=0.30,angle=0]{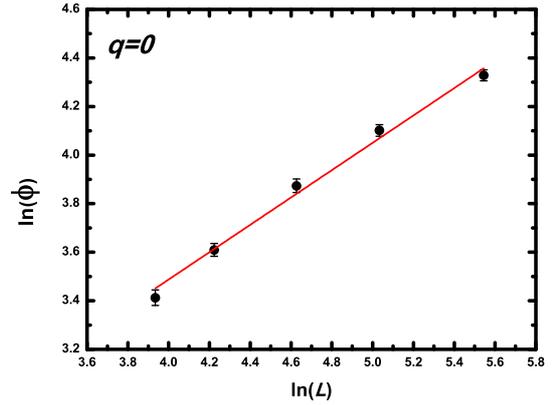}
 \caption{\label{fig:9} Log-log plot of $\phi$ (calculated at $p_{c}=0.275$) versus the linear size $L$ of the system for the case $q=0$. The red straight line is the best linear fit to the data ($\chi_{r}^2=2.08$ with a goodness-of-fit probability $Q(\chi_{r}^2)=10.02\%$).}
\end{figure}

\begin{figure}[!t]
 \centering
 \includegraphics*[scale=0.30,angle=0]{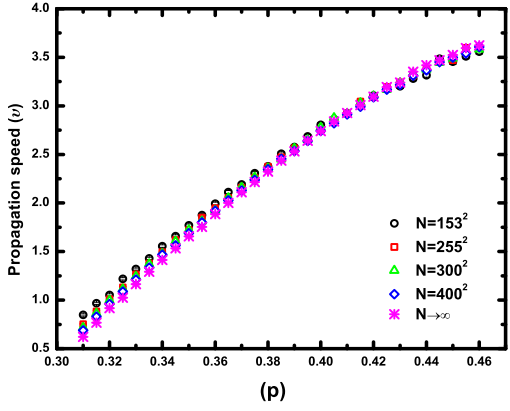}
 \caption{\label{fig:10} Propagation speed of the damage ($v$) on square lattices, for several lattice sizes along with an extrapolation of these data ($N\rightarrow\infty$). }
\end{figure}

\begin{figure}[!t]
 \centering
 \includegraphics*[scale=0.30,angle=0]{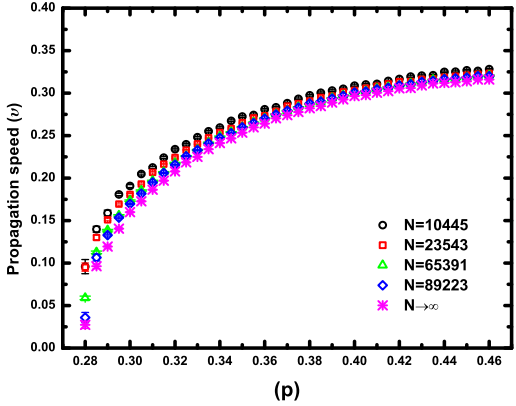}
 \caption{\label{fig:11} Propagation speed of the damage ($v$) on pure quasiperiodic lattices (case $q=0$), for several lattice sizes along with an extrapolation of these data ($N\rightarrow\infty$). }
\end{figure}

\begin{figure}[!t]
 \centering
 \includegraphics*[scale=0.30,angle=0]{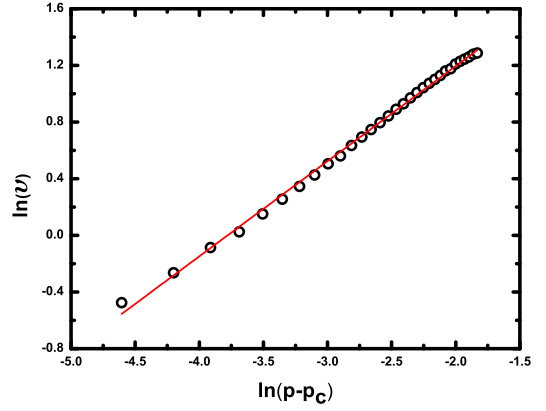}
 \caption{\label{fig:12} Log-log plot of $v$ versus $(p-p_{c})$ for the extrapolated data from Fig.~\ref{fig:10}. The red straight line is the best linear fit to Eq.~\ref{eq:11} on log-log scale.}
\end{figure}

\begin{figure}[!t]
 \includegraphics*[scale=0.30,angle=0]{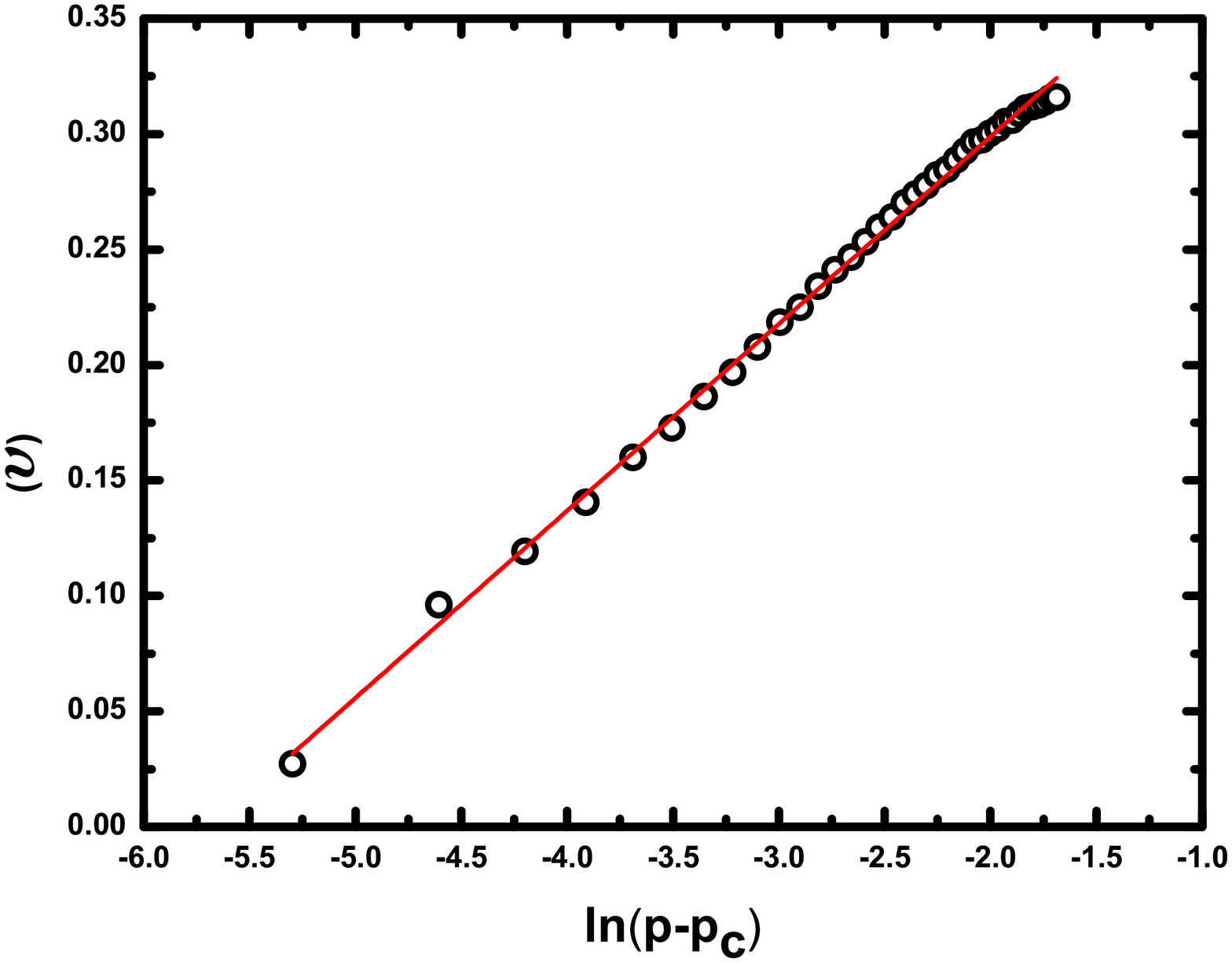}
 \caption{\label{fig:13} Semi-log plot of $v$ versus $(p-p_{c})$ for the extrapolated data from Fig.~\ref{fig:11}. The red straight line is the best linear fit to Eq.~\ref{eq:12}.}
\end{figure}

In order to determine the critical threshold parameter $p_{c}$ for the case $q=0$, we calculate the order parameter $\Psi$ in a wide range of values of $p$. Fig.~\ref{fig:2} shows the order parameter as a function of $p$ for three different lattice sizes ($N=10445, 23543$ and $65391$). Each data was averaged over $200$ different runs with an overall time of $5.0 \times 10^{3}$ time-steps for the system achieves its asymptotic regime. Looking at the inflection point of those curves, it can be noticed a typical second-order phase transition around $p_{c}=0.27(5)$. This value of critical point is below to that one found for regular lattices of 0.303~\cite{stauffer89,ferraz2007}. For the diluted case ($q\neq0$), we average over $600$ independent runs to get smoother curves of $\Psi$ against $p$. Fig.~\ref{fig:3} displays the phase transition on a lattice of $65391$ sites for three different value of the site dilution rate $q$. The damage spreading on such quasiperiodic lattices is quite sensitive to the removal of active sites. That is likely due to the existence of unstable regions (unstable cores)~\cite{klemm2005}. From Fig.~\ref{fig:3}, we see that even taking a small dilution rate $q=0.05$, the asymptotic damage mass is quite diminished and the value of the critical point on the lattice is shifted to $p_{c}=0.30(5)$. Even more dramatically, at $q=0.10$ and above we no longer observe a chaotic phase.

The statistical fluctuations play an important role for understanding the phase transition exhibited in this model. Figs.~\ref{fig:4} and \ref{fig:5} show the energy dispersion per spin $\Omega $ as a function of $p$ (Eq.~\ref{eq:5}) for three different lattice sizes, respectively, for the case $q=0$ and $q=0.05$. Remarkably, these curves display an abrupt vertical increase of the statical fluctuations on the energy values as defined by Eq.~\ref{eq:4} around their according critical points. However, it is interesting to observe that the maximum value of the dispersion is situated a little above $p_{c}$. Further, the energy dispersion diminishes in a different way when one moves away from the critical region towards the frozen region ($p<p_{c}$) than towards the chaotic region ($p>p_{c}$). To better characterize this asymmetric behaviour, we calculate the magnetic entropy $S$ (Eq.~\ref{eq:8}) by integrating the curves from Figs.~\ref{fig:4} and \ref{fig:5}. The absolute magnetic entropy $S$ as a function of the parameter $p$ considering a system of size $N=65391$ is shown in Fig.~\ref{fig:6} for both cases $q=0$ and $q=0.05$. In the frozen phase, the entropy exhibits a clear linear dependence on $p$ while in the chaotic phase it seems to increase in non-linear way as $p$ increases. A non-linear curve fitting to the data points above $p_{c}$ was performed for both cases $q=0$ and $q=0.05$. It is found that the best fits to the data are obtained by using the so-called monomolecular model, i.e.,
\begin{equation}\label{eq:9}
    S = A(1-\exp(-k(p -p_0 )),
\end{equation}
where the fit parameters $A$, $k$ and $p_{0}$ for both cases are given in Fig.~\ref{fig:6}. Such a change in the entropy behaviour is useful to characterize both phases with respect to the disorder degree present in these systems.

After the critical region and the critical parameter $p_{c}$ have been determined for each case and knowing that the final damage mass vanishes as $(p-p_{c})^\beta$ inside that region, we can estimate the critical exponents for both cases $q=0$ and $q=0.05$ by making a collapse of the data of $\Psi$ through the scaling law:
\begin{equation}\label{eq:9}
\psi(L,p)= L^{-\beta/ \nu}F((p-p_{c}) L^{1/\nu}),
\end{equation}
where $L=\sqrt{N}$ is the linear dimension of the lattice and $\nu$ is the exponent describing the divergence of the correlation length at $p_{c}$. Fig.~\ref{fig:7} shows the best data collapse for the case $q=0$ at $p_{c}=0.27(5)$. That data collapse was achieved taking $\beta=0.34\pm 0.05$ and $\nu=1.70\pm 0.09$. To support these estimates we also make a finite-size scaling analysis of the magnitude of the order parameter $\Psi^{\ast}$ at $p_{c}$. Fig.~\ref{fig:8} shows the log-log plot of $\Psi^{\ast}$ versus the linear size $L$ of the system. The slope of the linear fit to the data of $\Psi^{\ast}$ is $\beta/\nu=0.22\pm 0.01$. In this linear fit, the reduced chi-square $\chi_{r}^2$ was $1.14$ with a goodness-of-fit probability $Q(\chi_{r}^2)$ equals $33.4\%$, it means, the probability that $\chi_{r}^2$ would exceed the observed value, assuming that the underlying statistical model is correct. A typical confidence level is in accepting fits with $Q(\chi_{r}^2)>5\%$. The above estimate for the exponent $\beta/\nu$ yields a value of fractal dimension $D=d-\frac{\beta}{\nu}\approx1.78$ (where $d=2$ is the lattice dimension) in very good agreement with the one obtained to square lattices~\cite{stauffer89}.

Next, to estimate the correlation-length exponent $\nu$ we considered the power-law dependence of the logarithmic derivative of the order parameter ($\phi$) on the system size expressed by
\begin{equation}\label{eq:10}
\phi  = \frac{{d\ln(\Psi )}}{{dp}} \propto L^{1/\nu }.
\end{equation}
Fig.~\ref{fig:9} shows the log-log plot of $\phi$ (calculated at $p_{c}$) versus the linear size $L$ of the system. The slope of the linear fit to the data of $\phi$ is $1/\nu=0.58\pm 0.03$. The reduced chi-square $\chi_{r}^2$ was $2.08$ with a goodness-of-fit probability $Q(\chi_{r}^2)$ equals $10.02\%$. This value of $\nu=1.72\pm0.09$ is in good agreement with that one obtained via collapse of the data. Moreover, combining the last two estimated exponents: $\beta/\nu$ and $1/\nu$, we find that $\beta=0.38\pm0.03$ which is within the error bars in reasonable agreement with the value of $\beta$ obtained via collapse of the data. A similar analysis was also performed for the diluted case ($q=0.05$), for which we obtained $\beta=0.30\pm0.04$ and $\nu=1.70\pm0.05$. Therefore, based on the critical exponents hereby estimated we claim that on both pure ($q=0$) and diluted quasiperiodic lattices ($q=0.05$), the KCA model belongs to the same universality class than on square lattices.

The dynamics of the propagation of failures was studied for both periodic and quasiperiodic systems. Figs.~\ref{fig:10} and \ref{fig:11} display the propagation speed of the damage $v$ as a function of $p$, respectively, on square and on quasiperiodic lattices. In these figures, several lattice sizes are shown together with an extrapolation of these data for each value of $p$ considered.  As previously explained in the section \ref{sec:cp}, this extrapolation was achieved by analyzing how $v$ depends on the reciprocal of the lattice size ($1/N$) when one takes the limit $N\rightarrow\infty$. Fig.~\ref{fig:12} shows the log-log plot of $v$ versus $(p-p_{c})$ for the extrapolated data from Fig.~\ref{fig:10}, while Fig.~\ref{fig:13} shows the semi-log plot of $v$ versus $(p-p_{c})$ for the extrapolated data from Fig.~\ref{fig:11}. Thus, we can observe that the average propagation speed $v$ on square lattices follows a power law as:

\begin{equation}\label{eq:11}
v = v_{S}(p-p_c)^\alpha,
\end{equation}
where $v_{S}$ is a constant term and $\alpha\approx0.67$ is the critical exponent of the speed for the square lattice. While the average propagation speed $v$ on quasiperiodic lattices follows a logarithmic law as:

\begin{equation}\label{eq:12}
v=v_{Q}+\ln(p-p_c)^\alpha,
\end{equation}
where $v_{Q}$ is a constant term and $\alpha\approx0.08$ is the critical exponent of the speed for the quasiperiodic lattice. Therefore, our results lead us to conclude that quasiperiodic lattices are topologically more resistant than periodic lattices with respect to the propagation of failures generated by persistently disturbed sites.

\section{\label{sec:c} Summary and Conclusion}

In summary, we have employed the KCA model to study the breaking effects of the periodic translational symmetry on both the phase transition and the propagation dynamics of failures in quasiperiodic systems. These failures may mimic fractures, infectious disease spreading and infection by computer virus on lattices which possess short-range interactions but lacking a periodic translational symmetry as seen in quasicrystals.

Concerning the critical properties of the model, we have employed a finite-size scaling analysis to estimate the critical threshold parameter $p_{c}$ and its critical exponents $\beta/\nu$ and $1/\nu$ on both pure ($q=0$) and diluted quasiperiodic lattices ($q=0.05$).  In spite of the intrinsic complicate topology exhibits by these lattices, we claim that such systems are in the same universality class than on square lattices, although a phase transition takes place at a different critical threshold $p_{c}$. We have also defined an absolute magnetic entropy in order better to characterize the onset of chaos in these systems. It was seen that the value of the magnetic entropy linearly increases as the value of $p$ increases in the non-chaotic region; while in the chaotic region, it increases in a non-linear way as $p$ increases.

As for the robustness of the damage spreading, it was observed that such systems are quite sensitive when a small fraction of sites are disconnected from these lattices. In particular, for dilution rates of $q\geq0.10$, we did not find a chaotic phase anymore. In addition, the propagation dynamics of the damage was investigated by performing calculations of the propagation speed $v$ as a function of $p$ on both square (periodic) and quasiperiodic lattices. By using a data extrapolation procedure was found that the propagation speed of the damage on square lattices obeys a power law as $v\sim (p-p_{c})^{\alpha}$, whereas on quasiperiodic lattices it follows a logarithmic law as $v \sim \ln(p-p_{c})^\alpha$. For the square lattice, the estimated critical exponent of the speed ($\alpha$) was $0.67$, while for the quasiperiodic lattice was $0.08$. Therefore, we can conclude that quasiperiodic lattices are topologically more resistant than periodic lattices with respect to the propagation of failures when these are generated by persistently disturbed sites. Future work will concern numerical studies on $3D$ quasiperiodic lattices so that the more realistic structures found in alloys such as $Al$-$Fe$ and $Al$-$Mn$ can also be better investigated.

\section{Acknowledgements}
We wish to thank UFERSA for computational support.


\bibliographystyle{model1a-num-names}

\begin{thebibliography}{23}
\expandafter\ifx\csname natexlab\endcsname\relax\def\natexlab#1{#1}\fi
\providecommand{\bibinfo}[2]{#2}
\ifx\xfnm\relax \def\xfnm[#1]{\unskip,\space#1}\fi
\bibitem[{Kauffman(1969)}]{kauffman69}
\bibinfo{author}{S.~A. Kauffman}, \bibinfo{journal}{J. Theor. Biol.}
  \bibinfo{volume}{22} (\bibinfo{year}{1969}) \bibinfo{pages}{437}.
\bibitem[{Morelli and Zanetti(2001)}]{morelli2001}
\bibinfo{author}{L.~G. Morelli}, \bibinfo{author}{D.~H. Zanetti},
  \bibinfo{journal}{Phys. Rev. E} \bibinfo{volume}{63} (\bibinfo{year}{2001})
  \bibinfo{pages}{036204}.
\bibitem[{Bilke and Sjunnesson(2002)}]{bilke2002}
\bibinfo{author}{S.~Bilke}, \bibinfo{author}{F.~Sjunnesson},
  \bibinfo{journal}{Phys. Rev. E} \bibinfo{volume}{65} (\bibinfo{year}{2002})
  \bibinfo{pages}{016129}.
\bibitem[{Albert et~al.(2004)Albert, Albert, and Nakarado}]{reka2004}
\bibinfo{author}{R.~Albert}, \bibinfo{author}{I.~Albert},
  \bibinfo{author}{G.~L. Nakarado}, \bibinfo{journal}{Phys. Rev. E}
  \bibinfo{volume}{69} (\bibinfo{year}{2004}) \bibinfo{pages}{025103}.
\bibitem[{Luque and Sol{\'e}(1997)}]{luque97}
\bibinfo{author}{B.~Luque}, \bibinfo{author}{R.~V. Sol{\'e}},
  \bibinfo{journal}{Europhys. Lett.} \bibinfo{volume}{37}
  (\bibinfo{year}{1997}) \bibinfo{pages}{597}.
\bibitem[{Derrida and Stauffer(1986)}]{derrida86}
\bibinfo{author}{B.~Derrida}, \bibinfo{author}{D.~Stauffer},
  \bibinfo{journal}{Europhys. Lett.} \bibinfo{volume}{2} (\bibinfo{year}{1986})
  \bibinfo{pages}{739}.
\bibitem[{Stauffer(1987)}]{stauffer87}
\bibinfo{author}{D.~Stauffer}, \bibinfo{journal}{Philos. Mag. B}
  \bibinfo{volume}{56} (\bibinfo{year}{1987}) \bibinfo{pages}{901}.
\bibitem[{Arcangelis(1987)}]{arcangelis}
\bibinfo{author}{L.~Arcangelis}, \bibinfo{journal}{J. Phys. A}
  \bibinfo{volume}{20} (\bibinfo{year}{1987}) \bibinfo{pages}{369}.
\bibitem[{Stauffer(1994)}]{stauffer93}
\bibinfo{author}{D.~Stauffer}, \bibinfo{journal}{J. Stat. Phys}
  \bibinfo{volume}{74} (\bibinfo{year}{1994}) \bibinfo{pages}{1293}.
\bibitem[{Derrida and Pomeau(1986)}]{pomeau86}
\bibinfo{author}{B.~Derrida}, \bibinfo{author}{Y.~Pomeau},
  \bibinfo{journal}{Europhys. Lett.} \bibinfo{volume}{1} (\bibinfo{year}{1986})
  \bibinfo{pages}{45}.
\bibitem[{Ferraz and Herrmann(2007)}]{ferraz2007}
\bibinfo{author}{C.~H.~A. Ferraz}, \bibinfo{author}{H.~J. Herrmann},
  \bibinfo{journal}{Physica A} \bibinfo{volume}{373} (\bibinfo{year}{2007})
  \bibinfo{pages}{770}.
\bibitem[{Levine and Steinhardt(1986)}]{levine86}
\bibinfo{author}{D.~Levine}, \bibinfo{author}{P.~J. Steinhardt},
  \bibinfo{journal}{Phys. Rev. B} \bibinfo{volume}{34} (\bibinfo{year}{1986})
  \bibinfo{pages}{596}.
\bibitem[{Ferraz(2015)}]{ferraz2015}
\bibinfo{author}{C.~H.~A. Ferraz}, \bibinfo{journal}{Physica A}
  \bibinfo{volume}{440} (\bibinfo{year}{2015}) \bibinfo{pages}{90}.
\bibitem[{Stanley(1971)}]{stanley}
\bibinfo{author}{H.~E. Stanley}, \bibinfo{title}{Introduction to Phase
  Transitions and Critical Phenomena}, \bibinfo{publisher}{Oxford University
  Press}, \bibinfo{address}{Oxford}, \bibinfo{year}{1971}.
\bibitem[{Ferraz and Herrmann(2008)}]{ferraz2008}
\bibinfo{author}{C.~H.~A. Ferraz}, \bibinfo{author}{H.~J. Herrmann},
  \bibinfo{journal}{Physica A} \bibinfo{volume}{387} (\bibinfo{year}{2008})
  \bibinfo{pages}{5689}.
\bibitem[{Derrida and Flyvbjerg(1986)}]{flyvbjerg86}
\bibinfo{author}{B.~Derrida}, \bibinfo{author}{H.~Flyvbjerg},
  \bibinfo{journal}{J. Phys. A} \bibinfo{volume}{19} (\bibinfo{year}{1986})
  \bibinfo{pages}{1003}.
\bibitem[{Coniglio et~al.(1987)Coniglio, Stauffer, and Jan}]{coniglio87}
\bibinfo{author}{A.~Coniglio}, \bibinfo{author}{D.~Stauffer},
  \bibinfo{author}{N.~Jan}, \bibinfo{journal}{J. Phys. A} \bibinfo{volume}{20}
  (\bibinfo{year}{1987}) \bibinfo{pages}{1103}.
\bibitem[{Stephens et~al.(2013)Stephens, Mora, Tkacik, and Bialek}]{stephens}
\bibinfo{author}{G.~J. Stephens}, \bibinfo{author}{T.~Mora},
  \bibinfo{author}{G.~Tkacik}, \bibinfo{author}{W.~Bialek},
  \bibinfo{journal}{Phys. Rev. Lett.} \bibinfo{volume}{110}
  (\bibinfo{year}{2013}) \bibinfo{pages}{018701}.
\bibitem[{Wolfram(1984)}]{wolf}
\bibinfo{author}{S.~Wolfram}, \bibinfo{journal}{Physica D} \bibinfo{volume}{10}
  (\bibinfo{year}{1984}) \bibinfo{pages}{1}.
\bibitem[{Wolfram(1983)}]{wolf2}
\bibinfo{author}{S.~Wolfram}, \bibinfo{journal}{Rev. Mod. Phys.}
  \bibinfo{volume}{55} (\bibinfo{year}{1983}) \bibinfo{pages}{601}.
\bibitem[{Press et~al.(1996)Press, Teukolsky, Vetterling, Flannery, and
  Metcalf}]{recipes96}
\bibinfo{author}{W.~H. Press}, \bibinfo{author}{S.~A. Teukolsky},
  \bibinfo{author}{W.~T. Vetterling}, \bibinfo{author}{B.~P. Flannery},
  \bibinfo{author}{M.~Metcalf}, \bibinfo{title}{Numerical Recipes in Fortran
  90: The Art of Parallel Scientific Computing}, \bibinfo{publisher}{Cambridge
  University Press}, \bibinfo{address}{Cambridge}, p. \bibinfo{pages}{1053}.
\bibitem[{Stauffer(1989)}]{stauffer89}
\bibinfo{author}{D.~Stauffer}, \bibinfo{journal}{Physica D}
  \bibinfo{volume}{38} (\bibinfo{year}{1989}) \bibinfo{pages}{341}.
\bibitem[{Klemm and Bornholdt(2005)}]{klemm2005}
\bibinfo{author}{K.~Klemm}, \bibinfo{author}{S.~Bornholdt},
  \bibinfo{journal}{Phys. Rev. E} \bibinfo{volume}{72} (\bibinfo{year}{2005})
  \bibinfo{pages}{055101}.

\end{thebibliography}

\end{document}